\documentclass[prd,twocolumn,showpacs]{revtex4}
\usepackage{graphicx}

\newcommand{ \be}{\begin{equation}}
\newcommand{ \ee}{\end{equation}}
\newcommand{ \bea}{\begin{eqnarray}}
\newcommand{ \eea}{\end{eqnarray}}
\newcommand{ \mysmall}[1]{\scriptscriptstyle #1} 

\newcommand{ \eq}[1]{eq.~(\ref{eq:#1})}

\newcommand{ \bm}   {\boldmath}

\begin{document}


\title{Precise mass-dependent QED contributions to leptonic g-2 at order \bm
$\alpha^2$ and $\alpha^3$}


\author{M.~Passera}

\affiliation{Dipartimento di Fisica ``G.~Galilei'', Universit\`{a} 
        di Padova and \\ INFN, Sezione di Padova, I-35131, Padova, Italy} 


\begin{abstract} 

\noindent Improved values for the two- and three-loop mass-dependent
QED contributions to the anomalous magnetic moments of the electron,
muon, and $\tau$ lepton are presented. The Standard Model prediction
for the electron $(g-2)$ is compared with its most precise recent
measurement, providing a value of the fine-structure constant in
agreement with a recently published determination. For the $\tau$
lepton, differences with previously published results are found and
discussed. An updated value of the fine-structure constant is
presented in ``Note added after publication.''

\end{abstract}

\maketitle

\section{Introduction}
\label{sec:INTRO}

\noindent
The {\small QED} part of the anomalous magnetic moment $a_l\!=\!(g_l-2)/2$
of a charged lepton $l\!=\!e$, $\mu$ or $\tau$ arises from the subset of
Standard Model ({\small SM}) diagrams containing only leptons and
photons. For each of the three leptons $l$, of mass $m_l$, this
dimensionless quantity can be cast in the general form~\cite{KM90}
\be
    a_l^{\mysmall \rm QED} \!=\! A_1 + 
                   A_2 \!\left( \frac{m_l}{m_j} \right) + 
                   A_2 \!\left( \frac{m_l}{m_k} \right) + 
                   A_3 \!\left( \frac{m_l}{m_j},\frac{m_l}{m_k}\right),    
\label{eq:amuqedgeneral}
\ee
where $m_j$ and $m_k$ are the masses of the other two leptons. The term
$A_1$, arising from diagrams containing photons and leptons of only one
flavor, is mass and flavor independent.  In contrast, the terms $A_2$ and
$A_3$ are functions of the indicated mass ratios, and are generated by
graphs containing also leptons of flavors different from $l$. The
contribution of a lepton $j$ to $a_l^{\mysmall \rm QED}$ is suppressed by
$(m_l^2/m_j^2)$ if $m_j \!\gg\! m_l$, while it contains a logarithmic
enhancement factor $\ln(m_l/m_j)$ if $m_j \!\ll\! m_l$.  The muon
contribution to $a_e^{\mysmall \rm QED}$ is thus power suppressed by a
factor $(m_e^2/m_{\mu}^2) \sim 10^{-5}$; nonetheless, as we will discuss in
sec.~\ref{sec:ELECTRON}, this effect is much larger than the tiny
uncertainty very recently achieved in the measurement of
$a_e$~\cite{Gabrielse_g_2006}. On the contrary, the {\small QED} parts of
$a_{\mu,\tau}$ beyond one-loop are dominated by the mass-dependent terms.

The functions $A_i$ ($i\!=\!1,2,3$) can be expanded as power series in
$\alpha/\pi$ and computed order-by-order
\be
    A_i \!=\! A_i^{(2)}\!\left(\frac{\alpha}{\pi} \right)
    + A_i^{(4)}\!\left(\frac{\alpha}{\pi} \right)^{\!2}
    + A_i^{(6)}\!\left(\frac{\alpha}{\pi} \right)^{\!3}
    + A_i^{(8)}\!\left(\frac{\alpha}{\pi} \right)^{\!4} +\cdots.
\ee
Only one diagram is involved in the evaluation of the lowest-order
(first-order in $\alpha$, second-order in the electric charge) contribution
-- it provides the famous result by Schwinger $A_1^{(2)} \!=\!
1/2$~\cite{Sch48}. The mass-dependent coefficients $A_2$ and $A_3$ are of
higher order; the goal of this letter is to provide precise numerical values
for their $O(\alpha^2)$ and $O(\alpha^3)$ terms.  The relevance of the
results and the improvements with respect to earlier ones will be discussed
separately for each lepton.  All results were derived using the latest
{\small CODATA}~\cite{CODATA02} recommended mass ratios: $ m_e/m_{\mu} =
4.836 \, 331 \, 67 (13) \times 10^{-3}$, $ m_e/m_{\tau} = 2.875 \, 64 (47)
\times 10^{-4}$, $ m_{\mu}/m_e = 206.768 \, 2838 (54)$, $ m_{\mu}/m_{\tau} =
5.945 \, 92 (97) \times 10^{-2}$, $ m_{\tau}/m_e = 3477.48 (57)$, $
m_{\tau}/m_{\mu} = 16.8183 (27)$.
%
%
%
(The value for $m_{\tau}$ adopted by {\small CODATA} in ref.~\cite{CODATA02}
($m_{\tau}= 1776.99\, (29)$ MeV) is based on the 2002 {\small PDG}
result~\cite{PDG02}. This {\small PDG} result remains unchanged to
date~\cite{PDG04,PDG06}.)

\section{Electron}
\label{sec:ELECTRON}

\subsection{Two-loop contributions}
\label{subsec:2L}

\begin{figure}[h]
\begin{center}
\includegraphics[width=5.5cm]{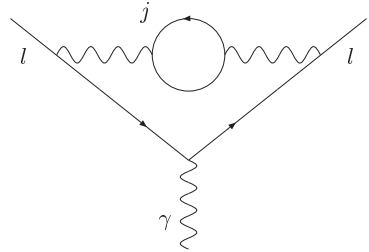}
\caption{The QED diagram generating the mass-dependent part of
    $a_l^{\mysmall \rm QED}$ in order $\alpha^2$.}
\label{fig:qed22}
\end{center}
\end{figure}

\noindent
Seven diagrams contribute to the fourth-order coefficient $A_1^{(4)}$, one
to $A_2^{(4)}(m_e/m_{\mu})$ and one to $A_2^{(4)}(m_e/m_{\tau})$. As there
are no two-loop diagrams contributing to $a_e^{\mysmall \rm QED}$ that
contain both virtual muons and taus, $A_3^{(4)}(m_e/m_{\mu},m_e/m_{\tau})
\!=\! 0$. The mass-independent coefficient has been known for almost fifty
years~\cite{So57-58-Pe57-58}:
\bea
    A_1^{(4)} &=& \frac{197}{144} + \frac{\pi^2}{12} 
                + \frac{3}{4}\zeta(3) - \frac{\pi^2}{2} \ln2 \nonumber \\
		&=& -0.328 \, 478 \, 965 \, 579 \, 193 \, 78 \ldots,
\label{eq:A14}
\eea
where $\zeta(s)$ is the Riemann zeta function of argument $s$. The
coefficient of the two-loop mass-dependent contribution to $a_l^{\mysmall
\rm QED}$, $A_2^{(4)}(1/x)$, with $x\!=\!m_j/m_l$, is generated by the
diagram in fig.~\ref{fig:qed22}, where $j$ is the virtual lepton in the
vacuum polarization subgraph.
This coefficient was first computed in the late 1950s for the muon
$g$$-$$2$ with $x \!=\! m_e/m_{\mu} \!\ll\! 1$, neglecting terms of
$O(x)$~\cite{SWP57}. The exact expression for $0\!<\!x\!<\!1$ was reported
by Elend in 1966~\cite{El66}. However, its numerical evaluation was
considered tricky because of large cancellations and difficulties in the
estimate of the accuracy of the results, so that common practice was to use
series expansions instead~\cite{Samuel91,Samuel93,CS99}. Taking advantage of
the properties of the dilogarithm ${\rm Li}_2(z)=-\!\int_0^z (dt/t)
\ln(1-t)$~\cite{Lewin}, the exact result was cast in~\cite{MP04} in a very
simple and compact analytic form, valid, contrary to the one in~\cite{El66},
also for $x \!\geq \! 1$ (the case relevant to $a_e^{\mysmall \rm QED}$ and
part of $a_{\mu}^{\mysmall \rm QED}$):
\bea 
     A_2^{(4)}(1/x)   &\!\!=\!\!& 
     -\frac{25}{36} - \frac{\ln x}{3} 
     +x^2 \left(4+3\ln x \right) + \! \frac{x}{2}\! \left(1-5 x^2\right) 
     \! \times 
     \nonumber \\     \times & & 
     \!\!\!\!\!\!\left[\frac{\pi^2}{2} 
       - \ln x \, \ln \left( \frac{1-x}{1+x} \right) 
       - {\rm Li}_2(x) + {\rm Li}_2(-x) \right] +
     \nonumber \\     +&& \!\!\!\!\!\!
     x^4 \left[ \frac{\pi^2}{3} -2\ln x \, \ln \left(\frac{1}{x}-x\right)
     -{\rm Li}_2(x^2)\right].
\label{eq:EA24}
\eea
For $x\!=\!1$, \eq{EA24} gives $A_2^{(4)}(1) \!=\! 119/36 - \pi^2/3$; of
course, this contribution is already part of $A_1^{(4)}$ in
\eq{A14}. Numerical evaluation of \eq{EA24} with the mass ratios given in
sec.~\ref{sec:INTRO} yields
\bea
     A_2^{(4)} (m_e/m_{\mu})  
     & = & 5.197 \, 386 \, 70 \, (28) \times 10^{-7} 
\label{eq:EA24m}
\\
     A_2^{(4)} (m_e/m_{\tau})  
     & = &  1.837 \, 62 \, (60) \times 10^{-9}, 
\label{eq:EA24t}
\eea
where the standard errors are only due to the uncertainties of the mass
ratios. The results of eqs.~(\ref{eq:EA24m}) and (\ref{eq:EA24t}) are equal
to those obtained with a series expansion in powers of $y$ and $\ln y$, with
$y\!  \ll \!1$~\cite{CODATA02}.

Adding up eqs.~(\ref{eq:A14}), (\ref{eq:EA24m}) and (\ref{eq:EA24t}) we get
the two-loop {\small QED} coefficient
\bea
    C^{(4)}_e &=& A_1^{(4)} + A_2^{(4)}(m_e/m_{\mu}) + 
               A_2^{(4)}(m_e/m_{\tau}) 
     \nonumber \\     &=& 
     - 0.328 \, 478 \, 444 \, 002 \, 90 \, (60).
\label{eq:EC2}
\eea 
The mass-dependent part of $C_e^{(4)}$ is small but not negligible, giving a
relative contribution to the theoretical prediction of the electron
$g$$-$$2$ of 2.4 parts per billion (ppb). This value is much larger than the
fabulous 0.7~ppb relative uncertainty very recently achieved in the
measurement of $a_e$~\cite{Gabrielse_g_2006}. The uncertainties in
$A_2^{(4)}(m_e/m_{\mu})$ and $A_2^{(4)}(m_e/m_{\tau})$ are dominated by the
latter and were added in quadrature. The resulting error $\delta C^{(4)}_e =
6 \! \times \! 10^{-13}$ leads to a totally negligible $O(10^{-18})$
uncertainty in the $a_e^{\mysmall \rm QED}$ prediction.

\subsection{Three-loop contributions}
\label{subsec:3L}

\noindent
More than one hundred diagrams are involved in the evaluation of the
three-loop (sixth-order) {\small QED} contribution. Their analytic
computation required approximately three decades, ending in the late 1990s.
The coefficient $A_1^{(6)}$ arises from 72 diagrams. Its exact expression,
mainly due to Remiddi and his collaborators, reads~\cite{Remiddi,LR96}:
\bea 
     A_1^{(6)} &=&  \frac{83}{72} \pi^2 \zeta(3) - \frac{215}{24}
     \zeta(5)  - \frac{239}{2160} \pi^4 + \frac{28259}{5184} +
     \nonumber \\ 
     && + \frac{139}{18} \zeta(3) - \frac{298}{9} \pi^2 \ln2 +
     \frac{17101}{810} \pi^2 +
     \nonumber \\ 
     && + \frac{100}{3} \left[{\rm Li}_4(1/2) + \frac{1}{24} \left( \ln^2
     2 -\pi^2 \right) \ln^2 2 \right]
     \nonumber \\ 
     &=& 1.181 \, 241 \, 456 \, 587 \ldots.
\label{eq:A16}
\eea
This value is in very good agreement with previous
results obtained with numerical methods~\cite{Ki90-95}.

The calculation of the exact expression for the coefficient
$A_2^{(6)}(m_l/m_j)$ for arbitrary values of the mass ratio $m_l/m_j$ was
completed in 1993 by Laporta and Remiddi~\cite{La93,LR93} (earlier works
include refs.~\cite{A26early}).  Let us focus on $a_e^{\mysmall \rm QED}$
($l\!=\!e$, $j\!=\!\mu$,$\tau$). This coefficient can be further split into
two parts: the first one, $A_2^{(6)}(m_l/m_j,\mbox{vac})$, receives
contributions from 36 diagrams containing either muon or $\tau$ vacuum
polarization loops~\cite{La93}, whereas the second one,
$A_2^{(6)}(m_l/m_j,\mbox{lbl})$, is due to 12 light-by-light scattering
diagrams with either muon or $\tau$ loops~\cite{LR93}. The exact expressions
for these coefficients are rather complicated, containing hundreds of
polylogarithmic functions up to fifth degree (for the light-by-light
diagrams) and complex arguments (for the vacuum polarization ones). Indeed,
they were too long to be listed in \cite{La93,LR93} (but were kindly
provided by their authors), although series expansions were given for
the cases of physical relevance. The exact expressions for the
light-by-light contributions also contain a few pentalogarithms in integral
form. We expressed these integrals in terms of harmonic polylogarithms
(introduced by Remiddi and Vermaseren in~\cite{HarmPol}), thus avoiding
their numerical integration.

The numerical evaluations of the exact expressions for
$A_2^{(6)}(m_l/m_j,\mbox{vac})$ and $A_2^{(6)}(m_l/m_j,\mbox{lbl})$ require
some care, as the presence of large cancellations makes them prone to
potentially large roundoff errors. For this reason, numerical evaluations
were carried out with {\tt Mathematica} codes employing exclusively
arbitrary-precision numbers, keeping track of precision at all
points~\cite{Mathematica}. Harmonic polylogarithms were implemented via the
{\tt Mathematica} package {\tt HPL}~\cite{Maitre05}.
Using the recommended mass ratios given in sec.~\ref{sec:INTRO}, we obtain
the following values:
\bea
     A_2^{(6)}(m_e/m_{\mu},\mbox{vac}) &=& -2.176 \, 840 \, 15 \,(11)
     \times 10^{-5}
\label{eq:EA26mvac}
     \\
     A_2^{(6)}(m_e/m_{\mu},\mbox{lbl}) &=& \:\: 1.439 \, 445 \, 989 \,(77)
     \times 10^{-5}
\label{eq:EA26mlbl}
     \\
     A_2^{(6)}(m_e/m_{\tau},\mbox{vac})&=& -1.167 \, 23 \,(36) 
     \times 10^{-7}
\label{eq:EA26tvac}
     \\
     A_2^{(6)}(m_e/m_{\tau},\mbox{lbl})&=& \:\: 5.0905 \,(17) 
     \times 10^{-8}.
\label{eq:EA26tlbl}
\eea
The sums of eqs.~(\ref{eq:EA26mvac})--(\ref{eq:EA26mlbl}) and 
eqs.~(\ref{eq:EA26tvac})--(\ref{eq:EA26tlbl}) are
\bea
     A_2^{(6)}(m_e/m_{\mu}) &=& -7.373 \, 941 \, 64 \,(29) 
     \times 10^{-6}
\label{eq:EA26m}
     \\
     A_2^{(6)}(m_e/m_{\tau})&=& -6.5819 \,(19) 
     \times 10^{-8}.
\label{eq:EA26t}
\eea
Equations (\ref{eq:EA26mvac})--(\ref{eq:EA26t}) provide the first evaluation
of the full analytic expressions for these coefficients with the {\small
CODATA} mass ratios of \cite{CODATA02}; they are almost identical to the
results $A_2^{(6)}(m_e/m_{\mu}) \!=\! -7.373 \, 941 \, 58 (28) \!\times\!
10^{-6}$ and $A_2^{(6)}(m_e/m_{\tau}) \!=\! -6.5819(19) \!\times\! 10^{-8}$
obtained in \cite{CODATA02} via the approximate series expansions in the
mass ratios. The small difference between $A_2^{(6)}(m_e/m_{\mu})$ of
\cite{CODATA02} and \eq{EA26m} mainly origins from the $O((m_e/m_{\mu})^6)$
term in the series expansion of $A_2^{(6)}(m_e/m_{\mu},\mbox{lbl})$; indeed,
due to its smallness, this term was neglected in the expansions \cite{LR93}
used in \cite{CODATA02}. Expanding the exact Laporta--Remiddi expression for
the sum of light-by-light and vacuum polarization contributions, for $r
\!=\!  m_l/m_{j} \!\ll\!  1$, we get
\bea 
      A_2^{(6)}(r) &=& \sum_{i=1}^4 r^{2i} f_{2i}(r) + 
      O\left(r^{10} \ln^2 \! r\right),
\label{eq:EA26mApprox}
      \\
     f_2(r) & = & \frac{23 \ln r}{135}+\frac{3 \zeta (3)}{2}
     -\frac{2 \pi^2}{45}-\frac{74957}{97200}, 
     \label{eq:f2} \\
     f_4(r) & = & -\frac{4337 \ln^2 \! r}{22680}+\frac{209891 \ln r}{476280}
     +\frac{1811 \zeta (3)}{2304} + \nonumber \\
     &-&  \!\frac{1919 \pi^2}{68040}-\frac{451205689}{533433600},
     \label{eq:f4} \\
     f_6(r) & = & -\frac{2807 \ln^2 \!r}{21600}+\frac{665641 \ln r}{2976750}
     +\frac{3077 \zeta (3)}{5760} + \nonumber \\
     &-& \!\frac{16967 \pi^2}{907200}- \frac{246800849221}{480090240000},
     \label{eq:f6} \\
     f_8(r) & = & \! -\frac{55163 \ln^2 \!r }{594000}  \!+\!
     \frac{24063509989 \ln r}{172889640000} \!+\! 
     \frac{9289 \zeta (3)}{23040}   \!+ \nonumber \\
     &-&  \!\frac{340019 \pi^2}{24948000}
     - \frac{896194260575549}{2396250410400000}.
     \label{eq:f8}
\eea
\noindent
The functions $f_2(r)$ and $f_4(r)$ coincide with the expansions provided
in~\cite{LR93}, and $f_6(r)$ agrees with the combination of parts
from~\cite{La93} (for the vacuum polarization contribution) and~\cite{KOPV}
(heavy-mass expansions for the light-by-light diagrams); $f_8(r)$ is new.
The value of $A_2^{(6)}(m_e/m_{\mu})$ obtained with
eq.~(\ref{eq:EA26mApprox}) perfectly agrees with that in
eq.~(\ref{eq:EA26m}) determined with the exact formulae. Indeed, their
difference is of $O(10^{-23})$, to be compared with the $O(10^{-13})$ error
$\delta A_2^{(6)}(m_e/m_{\mu})$ due to the present uncertainty of the ratio
$m_e/m_{\mu}$.
Therefore, it will be possible to compute $A_2^{(6)}(m_e/m_{\mu})$ with the
simple expansion in \eq{EA26mApprox} -- thus avoiding the complexities of
the exact expressions -- even if the precision of the ratio $m_e/m_{\mu}$
will improve in the future by orders of magnitude.

The contribution of the three-loop diagrams with both $\mu$ and $\tau$ loop
insertions in the photon propagator can be calculated numerically from the
integral expressions of ref.~\cite{Samuel91}. We get
\be
     A_3^{(6)}(m_e/m_{\mu},m_e/m_{\tau}) = 1.909 \, 45 \,(62) \times 10^{-13},
\label{eq:EA36}
\ee
a totally negligible $O(10^{-21})$ contribution to $a_{e}^{\mysmall \rm
QED}$.  Adding up eqs.~(\ref{eq:A16}), (\ref{eq:EA26m}), (\ref{eq:EA26t})
and (\ref{eq:EA36}) we obtain the three-loop {\small QED} coefficient
\be
    C^{(6)}_e    =   1.181 \, 234 \, 016 \, 827 \,(19).
\label{eq:EC3}
\ee
The relative contribution to $a_{e}^{\mysmall \rm QED}$ of the
mass-dependent part of this coefficient is $\sim 0.1$ ppb. This is smaller
than the present $\sim 0.7$ ppb experimental
uncertainty~\cite{Gabrielse_g_2006}.  The error $1.9 \! \times \! 10^{-11}$
in \eq{EC3} leads to a totally negligible $O(10^{-19})$ uncertainty in
$a_e^{\mysmall \rm QED}$.

\subsection{Determination of \boldmath $\alpha$ from the electron $g$$-$$2$
\unboldmath}
\label{subsec:ALPHA}

\noindent
Very recently, a new measurement of the electron anomalous magnetic moment
by Gabrielse and his collaborators achieved the fabulous relative
uncertainty of 0.66~ppb~\cite{Gabrielse_g_2006},
\be 
a_{e}^{\mysmall \rm EXP} = 115 \, 965 \, 218\, 0.85 \, (76) \times 10^{-12}.
\label{eq:aEEX}
\ee
This uncertainty is nearly six times smaller than that of the last
measurement of $a_e$ reported back in 1987, $a_{e}^{\mysmall \rm EXP} =
1159652188.3 (4.2) \!\times\!  10^{-12}$~\cite{UW87,CODATA02}. These two
measurements differ by 1.7 standard deviations.

The fine-structure constant $\alpha$ can be determined equating the
theoretical {\small SM} prediction of the electron $g$$-$$2$ with its
measured value
\be 
      a_{e}^{\mysmall \rm SM}(\alpha) = a_{e}^{\mysmall \rm EXP}.
\label{eq:SMvsEXP}
\ee
The {\small SM} prediction contains the {\small QED} contribution
$a_{e}^{\mysmall \rm QED}(\alpha)=\sum_{i=1}^5 C^{(2i)}_e (\alpha/\pi)^i$ 
(higher-order coefficients are assumed to be negligible), plus small weak
and hadronic loop effects:
$a_{e}^{\mysmall \rm SM}(\alpha) = a_{e}^{\mysmall \rm QED}(\alpha) +
 a_{e}^{\mysmall \rm EW}  + a_{e}^{\mysmall \rm HAD}$ 
(the dependence on $\alpha$ of any contribution other than $a_{e}^{\mysmall
\rm QED}$ is negligible). The electroweak contribution is~\cite{CODATA02}:
\be
     a_{e}^{\mysmall \rm EW} = 0.0297 \, (5) \times 10^{-12};
\label{eq:aEEW}
\ee
this precise value includes the two-loop contributions calculated in
ref.~\cite{CKM95}. The hadronic term is~\cite{CODATA02,AEHDR}:
\be
     a_{e}^{\mysmall \rm HAD} = 1.671 \, (19)\times 10^{-12}.
\label{eq:aEHD}
\ee
The latest value for the four-loop {\small QED} coefficient is
$C^{(8)}_e \!\!=\!-1.7283 (35)$~\cite{KN2005-4LE}.
Following the argument of~\cite{CODATA02}, we adopt the educated guess
$C^{(10)}_e  \!\!=\!0.0 (3.8)$ for the five-loop coefficient.
The errors $\delta C^{(8)}_e\!=\!0.0035$ and $\delta C^{(10)}_e \!=\! 3.8$
lead to an uncertainty of $0.1 \!\times\! 10^{-12}$ and $0.3 \!\times\!
10^{-12}$ in $a_e^{\mysmall \rm QED}$, respectively. Solving \eq{SMvsEXP}
with the new measured value of \eq{aEEX}, we obtain
\bea
\alpha^{-1} \, &=& \, 137.035 \, 999 \, 709 \, (12)\,(30)\,(2)\,(90) 
               \nonumber \\
               &=& \, 137.035 \, 999 \, 709 \, (96)\,[0.70~\mbox{ppb}]. 
\label{eq:alpha}
\eea
The first and second errors are due to the uncertainties of the four- and
five-loop {\small QED} coefficient $\delta C^{(8)}_e$ and $\delta
C^{(10)}_e$, respectively; the third one is caused by the tiny $\delta
a_{e}^{\mysmall \rm HAD}$, and the last one ($90 \!\times\! 10^{-9}$) is
from the experimental $\delta a_{e}^{\mysmall \rm EXP}$ in \eq{aEEX}.  The
uncertainty of the electroweak and two/three-loop {\small QED} contributions
are totally negligible at present.  The determination in \eq{alpha} is in
perfect agreement with the new result of ref.~\cite{Gabrielse_a_2006},
\be
\alpha^{-1} \, = \, 137.035 \, 999 \, 710 \, (96)
\label{eq:alphaG}
\ee
(also based on the new measurement of ref.~\cite{Gabrielse_g_2006}), whose
great precision represents the first significant improvement of this
fundamental constant in a decade. The totally negligible difference between
eqs.~(\ref{eq:alpha}) and (\ref{eq:alphaG}) is due to the rounded value
$a_{e}^{\mysmall \rm EW} = 0.030 \, (1) \times 10^{-12}$ \cite{CODATA02}
employed by the authors of ref.~\cite{Gabrielse_a_2006} instead of
\eq{aEEW}.

At present, the best determinations of $\alpha$ independent of the electron
$g$$-$$2$ are
\bea
\alpha^{-1} ({\rm Rb}) &=& 137.035 \, 998 \, 78 \, (91)\,[6.7~\mbox{ppb}], 
\label{eq:alphaRb}
\\
\alpha^{-1} ({\rm Cs}) &=& 137.036 \, 000 \, 0 \, (11) \,[8.0~\mbox{ppb}];
\label{eq:alphaCs}
\eea
they are less precise by roughly a factor of ten.  The value $\alpha^{-1}
({\rm Rb})$ was deduced from the measurement of the ratio $h/M_{\rm Rb}$
based on Bloch oscillations of Rb atoms in an optical lattice ($h$ is the
Planck constant and $M_{\rm Rb}$ is the mass of the Rb atom)~\cite{Rb2006},
while $\alpha^{-1} ({\rm Cs})$ was determined from the measurement of the
ratio $h/M_{\rm Cs}$ ($M_{\rm Cs}$ is the mass of the Cs atom) via cesium
recoil measurement techniques~\cite{Wicht2002,Cs2006}. These two
determinations of $\alpha$ also rely on the precisely known Rydberg constant
and relative atomic masses of the electron, Rb and Cs
atoms~\cite{CODATA02,RelMassesRbCs}.  The values of $\alpha$ in
eqs.~(\ref{eq:alphaRb}) and (\ref{eq:alphaCs}) are in good agreement with
the result of \eq{alpha}, differing from the latter by $-1.0$ and $+0.3$
standard deviations, respectively. This comparison provides a beautiful test
of the validity of {\small QED}. It also probes for possible electron
substructure~\cite{Gabrielse_a_2006}.

\section{Tau}
\label{sec:TAU}

\noindent
The two-loop mass-dependent {\small QED} contributions to the anomalous
magnetic moment of the $\tau$, obtained by direct evaluation of the exact
formula in \eq{EA24}, are
\bea
     A_2^{(4)}(m_{\tau}/m_e)      & = & 2.024 \, 284 \, (55),  
\label{eq:TA24e}
\\
     A_2^{(4)}(m_{\tau}/m_{\mu}) & = & 0.361 \, 652 \, (38).
\label{eq:TA24m}
\eea
These two values are very similar to those computed via a dispersive
integral in ref.~\cite{Narison01} (which, however, contain no estimates of
the uncertainties). Equations~(\ref{eq:TA24e}) and (\ref{eq:TA24m}) are also
in agreement (but more accurate) with those of ref.~\cite{Samuel93}.
Adding up eqs.~(\ref{eq:A14}), (\ref{eq:TA24e}) and (\ref{eq:TA24m}) we
get
\be
       C_{\tau}^{(4)} = 2.057 \, 457 \, (93)
\label{eq:TC2}
\ee
(note that the uncertainties in $m_{\tau}/m_e$ and $m_{\tau}/m_{\mu}$ are
correlated). The resulting error $9.3 \times \!  10^{-5}$ leads to a $5 \!
\times \! 10^{-10}$ uncertainty in $a_{\tau}^{\mysmall \rm QED}$.

We computed the three-loop mass-dependent contributions by direct numerical
evaluation of the exact analytic expressions (see
sec.~\ref{subsec:3L}). The results are:
\bea
     A_2^{(6)}(m_{\tau}/m_{e},\mbox{vac}) &=&  7.256 \, 99 \,(41)
\label{eq:TA26evac}
     \\
     A_2^{(6)}(m_{\tau}/m_{e},\mbox{lbl}) &=&  39.1351 \,(11)
\label{eq:TA26elbl}
     \\
     A_2^{(6)}(m_{\tau}/m_{\mu},\mbox{vac})&=& -0.023 \, 554 \,(51) 
\label{eq:TA26mvac}
     \\
     A_2^{(6)}(m_{\tau}/m_{\mu},\mbox{lbl})&=&  7.033 \, 76 \,(71).
\label{eq:TA26mlbl}
\eea
Employing the approximate series expansions (see sec.~\ref{subsec:3L})
we obtain almost identical values:
7.25699(41),
39.1351(11),
$-$0.023564(51), 
7.03375(71).
The estimates of ref.~\cite{Narison01} were: 10.0002, 39.5217, 2.9340, and
4.4412 (no error estimates were provided), respectively; they are at
variance with our results, eqs.~(\ref{eq:TA26evac})--(\ref{eq:TA26mlbl}),
derived from the exact analytic expressions. The estimates of
ref.~\cite{Samuel_tau} compare slightly better: 7.2670, 39.6, $-0.1222$,
4.47 (no errors provided). In the specific case of
$A_2^{(6)}(m_{\tau}/m_{\mu},\mbox{lbl})$ it's easy to check that the values
of refs.~\cite{Narison01,Samuel_tau} differ from \eq{TA26mlbl} because their
derivations did not include terms of $O(m_{\mu}/m_{\tau})$, which turn out
to be unexpectedly large.  The sums of
eqs.~(\ref{eq:TA26evac})--(\ref{eq:TA26elbl}) and
(\ref{eq:TA26mvac})--(\ref{eq:TA26mlbl}) are
\bea
     A_2^{(6)}(m_{\tau}/m_e) &=& 46.3921 \,(15), 
\label{eq:TA26e}
     \\
     A_2^{(6)}(m_{\tau}/m_{\mu})&=& 7.010 \, 21 \,(76).
\label{eq:TA26m}
\eea
The contribution of the three-loop diagrams with both electron- and
muon-loop insertions in the photon propagator can be calculated numerically
from the integral expressions of \cite{Samuel91}. We get
\be
     A_3^{(6)}(m_{\tau}/m_e,m_{\tau}/m_{\mu}) = 3.347 \, 97 \,(41).
\label{eq:TA36}
\ee
This value disagrees with the results of refs.~\cite{Samuel_tau}
(1.679) and \cite{Narison01} (2.75316). Combining the three-loop results of
eqs.~(\ref{eq:A16}), (\ref{eq:TA26e}), (\ref{eq:TA26m}) and (\ref{eq:TA36})
we find the sixth-order {\small QED} coefficient 
\be
      C_{\tau}^{(6)}  =  57.9315 \,(27).
\label{eq:TC3}
\ee
The error $2.7 \!\times\! 10^{-3}$ induces a $3 \! \times \!  10^{-11}$
uncertainty in $a_{\tau}^{\mysmall \rm QED}$. The order of magnitude of the
three-loop contribution to $a_{\tau}^{\mysmall \rm QED}$, dominated by the
mass-dependent terms, is comparable to that of electroweak and hadronic
effects.

Adding up all the above contributions and using the new value
$\alpha^{-1} \, = \,  137.035 \, 999 \, 710 \, (96)$ \cite{Gabrielse_a_2006}
(or the value derived in \eq{alpha},
$\alpha^{-1} \, = \,  137.035 \, 999 \, 709 \, (96)$
-- the difference is totally negligible)
we obtain the total {\small QED} contribution to the $g$$-$$2$ of the
$\tau$:
\be
    a_{\tau}^{\mysmall \rm QED} =
    117 \, 324 \, (2) \times 10^{-8}.
\label{eq:TQED}
\ee
The error $\delta a_{\tau}^{\mysmall \rm QED}$ is the uncertainty
$\delta C_{\tau}^{(8)}(\alpha/\pi)^4 \sim \pi^2 \ln^2(m_{\tau}/m_e)
(\alpha/\pi)^4 \sim 2\times 10^{-8}$
which we assigned to $a_{\tau}^{\mysmall \rm QED}$ for uncalculated
four-loop contributions. As we mentioned earlier, the errors due to the
uncertainties of the $O(\alpha^2)$ and $O(\alpha^3)$ terms are negligible.
The error induced by the uncertainty of $\alpha$ is only $8 \!\times\!
10^{-13}$ (and thus totally negligible). 

The $g$$-$$2$ of the $\tau$ is a very interesting observable, even if the
short lifetime of this lepton makes its measurement very difficult at
present. The possibility to improve the recent experimental
bounds~\cite{PDG06} is certainly not excluded.

\section{Muon}
\label{sec:MUON}

\noindent 
This final section reports the results relevant to $a_{\mu}^{\mysmall \rm
QED}$ (see \cite{g-2review} for recent reviews of the entire {\small SM}
prediction). Some of them were already presented in~\cite{MP04}. The
two-loop contributions are
\bea
     A_2^{(4)}(m_{\mu}/m_e)      & = & 1.094\,258\,3111 \, (84),  
\label{eq:MA24e}
\\
     A_2^{(4)}(m_{\mu}/m_{\tau}) & = & 0.000\,078\,064 \, (25).
\label{eq:MA24tau}
\eea
The sum of eqs.~(\ref{eq:A14}), (\ref{eq:MA24e}) and (\ref{eq:MA24tau})
provides the coefficient $C_{\mu}^{(4)} \!=\! 0.765 \, 857 \, 410 \,(27)$.
The value $\delta C_{\mu}^{(4)} \!=\! 2.7 \! \times \! 10^{-8}$ was obtained
adding in quadrature the errors in eqs.~(\ref{eq:MA24e}) and
(\ref{eq:MA24tau}). It produces a tiny $1.4 \!  \times \! 10^{-13}$
uncertainty in $a_{\mu}^{\mysmall \rm QED}$. The three-loop contributions
are
\bea
     A_2^{(6)}(m_{\mu}/m_e,\mbox{vac}) &=&  \:\: 1.920\, 455 \, 130 \, (33)
\label{eq:MA26evac}
     \\
     A_2^{(6)}(m_{\mu}/m_e,\mbox{lbl}) &=&      20.947 \, 924 \, 89\,(16)
\label{eq:MA26elbl}
     \\
     A_2^{(6)}(m_{\mu}/m_{\tau},\mbox{vac})&=& \!\! -0.001\,782\,33 \, (48)  
\label{eq:MA26tvac}
     \\
     A_2^{(6)}(m_{\mu}/m_{\tau},\mbox{lbl})&=&\,\:\,0.002\,142\,83\, (69)
\label{eq:MA26tlbl}
     \\
     A_2^{(6)}(m_{\mu}/m_e) &=&      22.868 \, 380 \, 02\,(20)
\label{eq:MA26e}
     \\
     A_2^{(6)}(m_{\mu}/m_{\tau})&=&\,\:\,0.000\, 360 \, 51\, (21).
\label{eq:MA26t}
\eea
The analytic calculation of the three-loop diagrams with both electron and
$\tau$ loop insertions in the photon propagator became available in
1999~\cite{CS99} and was confirmed more recently \cite{deRafael05}. This
analytic result yields the numerical value~\cite{MP04}
\be
     A_3^{(6)}(m_{\mu}/m_e,m_{\mu}/m_{\tau}) = 0.000 \, 527 \, 66
     \,(17),
\label{eq:MA36}
\ee
providing a small $0.7 \times 10^{-11}$ contribution to $a_{\mu}^{\mysmall
\rm QED}$. The error $1.7 \times 10^{-7}$ is caused by the uncertainty of
the ratio $m_{\mu}/m_{\tau}$. Combining the three-loop results of
eqs.~(\ref{eq:A16}), (\ref{eq:MA26e}), (\ref{eq:MA26t}) and (\ref{eq:MA36})
we get the three-loop coefficient $C_{\mu}^{(6)} = 24.050 \, 509 \,64 \,
(43)$. The error $4.3 \times 10^{-7}$ induces a negligible $O(10^{-14})$
uncertainty in $a_{\mu}^{\mysmall \rm QED}$.

Adding the four-loop and leading five-loop contributions computed by
Kinoshita and Nio,
$C_{\mu}^{(8)} = 130.9916 \, (80)$~\cite{KN2005-4LE,KN2004-4LM} and
$C_{\mu}^{(10)} = 663 \, (20)$~\cite{KN2006-5LM} (estimates obtained with
the renormalization-group method agree with this five-loop
result~\cite{Kataev}),
and using the new value
$\alpha^{-1} \, = \,  137.035 \, 999 \, 710 \, (96)$ \cite{Gabrielse_a_2006}
(or the value derived in \eq{alpha},
$\alpha^{-1} \, = \,  137.035 \, 999 \, 709 \, (96)$
-- the difference is negligible)
we get the new total {\small QED} contribution to the muon $g$$-$$2$, 
\be
     a_{\mu}^{\mysmall \rm QED} =
     116 \, 584 \, 718.09 \, (14)\,(08)  \times 10^{-11}.
\label{eq:MaQED}
\ee
The first error is determined by the uncertainties of the {\small QED}
coefficients (dominated by the five-loop one, $\delta C_{\mu}^{(10)}\!=
20$), while the second is caused by the tiny uncertainty
$\delta\alpha$. Equation~(\ref{eq:MaQED}) is in good agreement with the
recent value $a_{\mu}^{\mysmall \rm QED} = 116 \, 584 \, 717.62 \,
(14)\,(85) \times 10^{-11}$~\cite{KN2006-5LM}, and the uncertainty due to
$\delta\alpha$ is strongly reduced.

\subsection*{Note added after publication}
\label{sec:NOTE}

The value of the mass-independent eighth-order {\small QED}
coefficient $A_1^{(8)}$ has been recently revised by Kinoshita and
collaborators~\cite{KN07}, inducing the following updates.

The revision of $A_1^{(8)}$ (and, consequently, of the total four-loop
{\small QED} coefficient $C^{(8)}_e \!\!  \simeq \!\!  A_1^{(8)}$),
from
$-1.7283 (35)$~\cite{KN2005-4LE}
to 
$-1.9144 (35)$~\cite{KN07}, 
induces a shift in the value of $\alpha$ from \eq{alpha} to
\bea
\alpha^{-1} \, &=& \, 137.035 \, 999 \, 068 \, (12)\,(30)\,(2)\,(90) 
               \nonumber \\
               &=& \, 137.035 \, 999 \, 068 \, (96)\,[0.70~\mbox{ppb}]. 
\label{eq:newalpha}
\eea
This new value is still in good agreement with those in
eqs.~(\ref{eq:alphaRb}) and (\ref{eq:alphaCs}), which are less precise
by roughly a factor of ten and differ from the value in \eq{newalpha}
by $-0.3$ and $+0.8$ standard deviations, respectively.

The total {\small QED} contribution to the muon $g$$-$$2$ shifts from
the value in \eq{MaQED} to
\be
     a_{\mu}^{\mysmall \rm QED} =
     116 \, 584 \, 718.10 \, (14)\,(08)  \times 10^{-11}.
\label{eq:newMaQED}
\ee
This tiny variation is only of $O((\alpha/\pi)^5)$. Indeed, the
$O((\alpha/\pi)^4)$ shift in $a_{\mu}^{\mysmall \rm QED}$ due to the
revision of $A_1^{(8)}$ (and, consequently, of the total four-loop
{\small QED} coefficient, now standing at
$C_{\mu}^{(8)} = 130.8055 \, (80)$)
is compensated by the $O((\alpha/\pi)^4)$ change in the value of
$\alpha$ determined from the electron $g$$-$$2$.

The shift in the value of $\alpha$ from \eq{alpha} to \eq{newalpha}
induces no appreciable variation in the total {\small QED}
contribution to the $g$$-$$2$ of the $\tau$ lepton.

\begin{acknowledgments}
\noindent 
I wish to thank A.~Ferroglia, M.~Giacomini, T.~Kinoshita, D.~Ma\^{\i}tre,
P.~Minkowski, and P.J.~Mohr for useful discussions and correspondence.
\end{acknowledgments}


\end{document}